\documentclass[12pt]{iopart}
\usepackage{graphicx}
\begin{document}

\title[Single-crystal growth of La$_{0.34}$Na$_{0.66}$Fe$_2$As$_2$]{Single-crystal growth of iron-based superconductor La$_{0.34}$Na$_{0.66}$Fe$_2$As$_2$}

\author{Yanhong Gu$^{1,2}$, Jia-Ou Wang$^3$, Xiaoyan Ma$^{1,2}$, Huiqian Luo$^1$, Youguo Shi$^1$, Shiliang Li$^{1,2,4}$}
\address{$^1$ Beijing National Laboratory for Condensed Matter Physics, Institute of Physics, Chinese Academy of Sciences, Beijing 100190, China}
\address{$^2$ School of Physical Sciences, University of Chinese Academy of Sciences, Beijing 100190, China}
\address{$^3$ Being Synchrotron Radiation Facility, Institute of High Energy Physics, Beijing 100049, China}
\address{$^4$ Collaborative Innovation Center of Quantum Matter, Beijing 100190, China}
\ead{slli@iphy.ac.cn}

\begin{abstract}
We report single-crystal growth of La$_{0.34}$Na$_{0.66}$Fe$_2$As$_2$ iron-based superconductor with the size of several millimeters. The samples were accidentally obtained in trying to grow LaFeAsO$_{1-y}$F$_y$  ( $y \geq$ 0.8 ) single crystals with NaAs and NaF flux. The sample shows both antiferromagnetic and structural transitions at 106 K. The superconducting transition temperature is about 27 K with the superconducting anisotropy of about 1.9. These values and the temperature dependence of the Hall coefficient suggest that La$_{0.34}$Na$_{0.66}$Fe$_2$As$_2$ belongs to hole-doped "122" families of iron pnictides with the doping level slightly lower than optimal doping. Compared with previous reports on the polycrystalline samples, our results suggest that either $T_c$ of the this system can be further increased, or the superconducting dome may not be well formed in this system. In either case, the La$_{0.5-x}$Na$_{0.5+x}$Fe$_2$As$_2$ system provides a new platform to study the antiferromagnetic and superconducting properties of iron-based superconductors.
\end{abstract}

%
\vspace{2pc}

\section{Introduction}

Among various families of iron-based superconductors, the so-called "122" iron pnictides have attracted many interests due to the available of high-quality large-size single crystals \cite{HosonoH15,LuoX15,DaiP15,YiM17}. The parent compounds of "122" materials are typically in the form of A$_e$Fe$_2$As$_2$ (A$_e$ = alkaline earth metal, e.g., Ca, Sr, Ba), showing both antiferromagnetic (AF) and structural transitions \cite{HuangQ08,KrellnerC08,RonningF08}. Superconductivity can be achieved through either hole or electron doping by substituting A$_e$ with alkali metal elements (e.g. Na, K) \cite{RotterM08,GokoT09,ZhaoK11,ShinoharaN15} and Fe with transition metals (e.g., Co, Ni) \cite{SefatAS08,LiJL09,NiN09} respectively. The phase diagrams of these materials show very significant electron-hole asymmetry \cite{FangL09,AvciS12,NeupaneM11}, i.e., superconductivity is achieved and optimized at different concentrations for electron and hole carriers. While this asymmetry may have underlying physics \cite{FangL09,AvciS12,NeupaneM11,XuG09,DaiP12,GuY17}, it is hard to rule out the possibility that it may come from the fact that electron and hole doping are obtained by substituting elements at different sites, especially considering that the electron doping involves the change of FeAs layers.

In growing LaFeAsO single crystals, it has been found that the single crystal of La$_{0.4}$Na$_{0.6}$Fe$_2$As$_2$ can be accidentally obtained \cite{YanJQ15}. The latter has the same crystal structure as A$_e$Fe$_2$As$_2$ with both La and Na occupy at the $A_e$ site. It shows first-order AF and structural transitions at $T_N = T_s$ = 125 K, which is very similar to underdoped hole-doped Ba$_{1-x}$K$_x$Fe$_2$As$_2$ \cite{AvciS12}. Filamentary superconductivity is found when the sample is immersed in water, most likely due to the effect on its surface. It has been thus proposed that the La$_{0.5-x}$Na$_{0.5+x}$Fe$_2$As$_2$ may provide a unique platform to study the electron-hole asymmetry without disturbing the FeAs layer since both types of carriers may be introduced from the "parent" compound of La$_{0.5}$Na$_{0.5}$Fe$_2$As$_2$ \cite{YanJQ15} by changing the ratio of La and Na. Recently, the hole-doped polycrystalline samples for nominal $x$ from 0 to 0.35 have been successfully synthesized \cite{IyoA18}. The phase diagram is similar to those of sodium doped A$_e$Fe$_2$As$_2$, showing that $x$ for the parent compound of this system is 0, and superconductivity can be achieved for $x \geq$ 0.15 with the maximum superconducting transition temperature $T_c$ of about 27 K for $x$ above 0.3. Unfortunately, the electron-doped side, i.e., $x <$ 0, cannot be synthesized. Moreover, the physical properties of this system have not be studied in details due to the lack of single crystals.

In this paper, we report the single-crystal growth of La$_{0.34}$Na$_{0.66}$Fe$_2$As$_2$ with $T_c$ = 27 K. Both the resistivity and hall measurements reveal that the AF and structural transitions happen at 106 K, suggesting that the doping level is slightly lower than optimal doping level \cite{ShenB11,AswarthamS12,OhgushiK12,LiuY14}. The superconducting anisotropy ratio $\Gamma$ for the upper critical fields is about 1.9, which is similar to those in the hole-doped "122" systems \cite{NiN08,WangZS08,SunDL09,HaberkornN11}. Surprisingly, the $T_c$ of our samples is already the same as the maximum value reported in polycrystalline samples \cite{IyoA18}, suggesting that either the value of $T_c$ can be further enhanced, or the superconducting dome may be absent in this system.

\section{Experiments}

As reported previously, the La$_{0.34}$Na$_{0.66}$Fe$_2$As$_2$ single crystals were obtained accidentally \cite{YanJQ15}. Therefore, the following procedure is actually used to grow the LaFeAsO$_{1-x}$F$_x$ single crystals \cite{YanJQ09,GuY18}. The starting materials were La ( 99.7\% ), As ( 99.99\% ), Fe ( 99.998\% ), Fe$_2$O$_3$ ( 99.998\% ), NaF ( 99.99\% ) and FeF$_2$ ( 98\% ). LaAs powders were prepared by reacting La chips and As chips at $500^{\circ}C$ for 15 hours and then $850^{\circ}C$ for 15 hours. Powders of LaAs, Fe$_2$O$_3$, Fe and FeF$_2$ were mixed together according to the ratio La: Fe: As: O: F =1: 1: 1: 1-$x$: $x$ with $x$ ranging from 0.8 to 0.98, and then pressed to pellets. The pellets were put into an Al$_2$O$_3$ crucible and then sealed into an evacuated quartz tube, which was heated at 1150$^{\circ}C$ for 60 hours and then slowly cooled down to room temperature. Flux NaAs were prepared by Na chunk and As chips in an Al$_2$O$_3$ crucible sealed into an evacuated quartz tube, which was heated at $400^{\circ}C$ for 20 hours and then cooled down to room temperature with intermediate grindings. The nominal LaFeAsO$_{1-x}$F$_x$ pellets, NaAs and NaF were grounded together with a molar ratio of 1:17:11 and sealed into a Ta tube under atmosphere of argon. To avoid the oxidation at high temperature, the Ta tube was sealed into an evacuated quartz tube and heated at $1150^{\circ}C$ for 10 hours, then slowly cooled down to 700$^{\circ}C$ at a speed 2$^{\circ}C/$h followed by a fast cooling down to room temperature. Plate-like single crystals were obtained by dissolving the final productions of the above procedure in water.

\begin{figure}[tbp]
\includegraphics[width=\columnwidth]{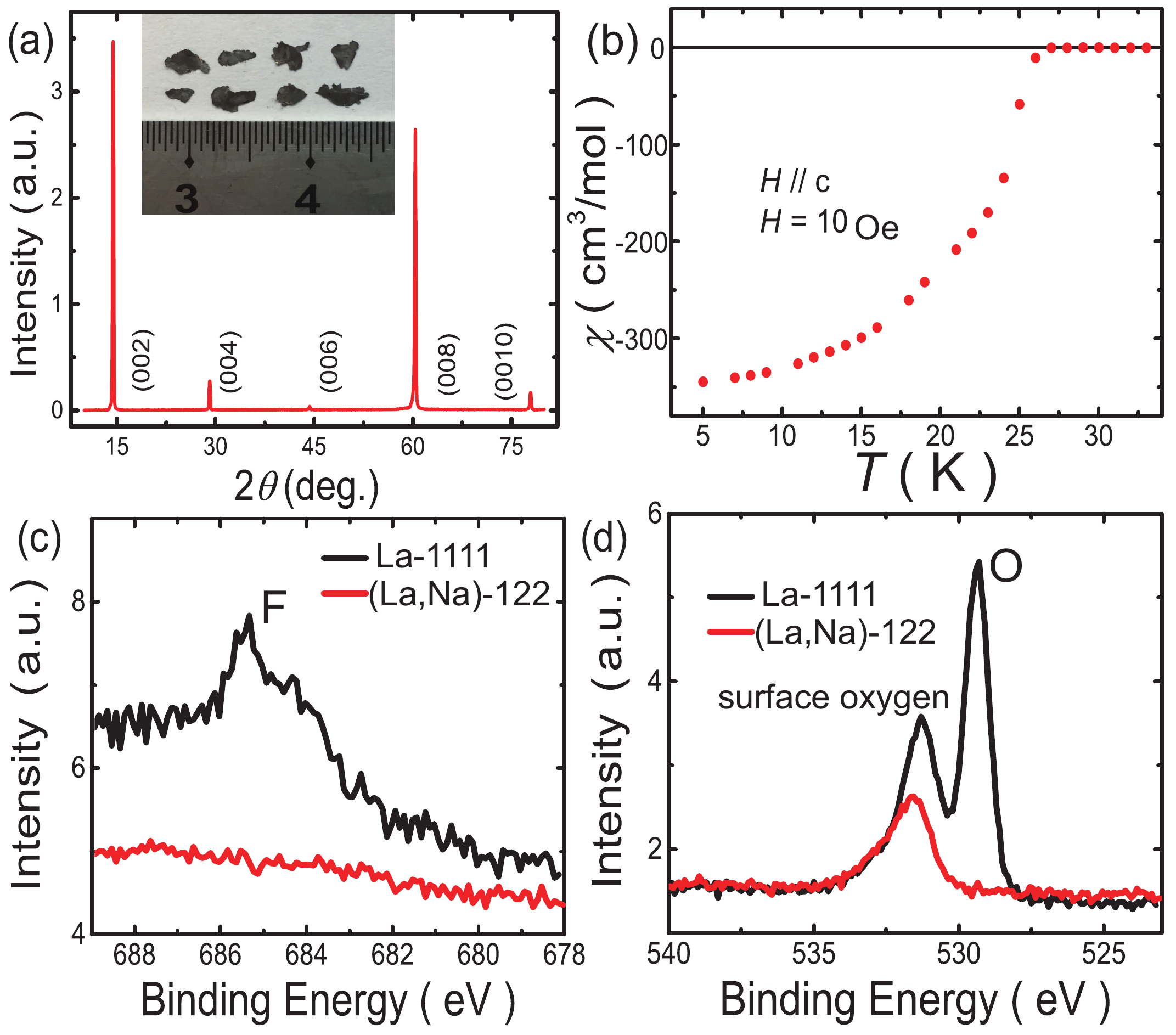}
\caption{(a) X-ray diffraction pattern along c-axis direction. The inset shows photo of the single crystals of La$_{0.34}$Na$_{0.66}$Fe$_2$As$_2$, the unit of numbers on the ruler is centimeter. (b) Temperature dependence of the DC-magnetic susceptibility at 10 Oe with field direction parallel to c-axis. (c) \& (d) XPS core level spectra for LaFeAsO$_{0.74}$F$_{0.26}$ (La-1111) and La$_{0.34}$Na$_{0.66}$Fe$_2$As$_2$ ((La,Na)-122) at the binding energy for fluorine and oxygen elements, respectively.}
\end{figure}

The composition and lattice parameters are determined by the single-crystal X-ray diffraction (SC-XRD), the energy dispersive X-ray (EDX) and the inductively coupled plasma (ICP) analysis. The analysis of the EDX spectrum is obtained by averaging from the measurements on different areas of several single crystals. The X-ray photoelectron spectroscopy(XPS) was measured on 4B9B beamline - Photoelectron Spectroscopy Station at Beijing Synchrotron Radiation Facility. The resistivity and Hall resistivity were measured by the standard four-probe method in a Physical Property Measurement System (PPMS, Quantum Design). The latter was obtained by averaging the values at positive and negative fields to avoid the effect of magnetoresistivity. The DC magnetic susceptibility was measured in a Magnetic Property Measurement System (MPMS, Quantum Design).

\section{Results}

The inset of Fig. 1(a) shows the photo of the as-grown single crystals, which are plate-like with the in-plane size of several millimeters and the thickness of several micrometers. Figure 1 shows the powder X-ray diffraction result on the in-plane of the crystal, which only shows sharp (0,0,L) peaks similar to those reported in Ref. \cite{YanJQ15}. The magnetic susceptibility measurement shows clear diamagnetic signal and gives a $T_c$ of 27 $\pm$ 0.5 K, as shown in Fig. 1(b). Both the EDX and ICP measurements suggest that there is neither fluorine nor oxygen element in the samples and the ratio of (La+Na):Fe:As is close to 1:2:2, suggesting that La$_{0.5-x}$Na$_{0.5+x}$Fe$_2$As$_2$ has been successfully grown. We have also measured the XPS core level spectra and compared the results with those of LaFeAsO$_{0.74}$F$_{0.26}$ \cite{note1}, as shown in Fig. 1(c) and 1(d). The oxygen peaks between 531 eV to 532 eV are from the oxygen contamination on the surfaces. The peaks at about 685 eV and 529 eV are for the fluorine and oxygen core levels, respectively. Apparently, the samples studied here contain neither fluorine nor oxygen.

\begin{figure}[tbp]
\includegraphics[width=\columnwidth]{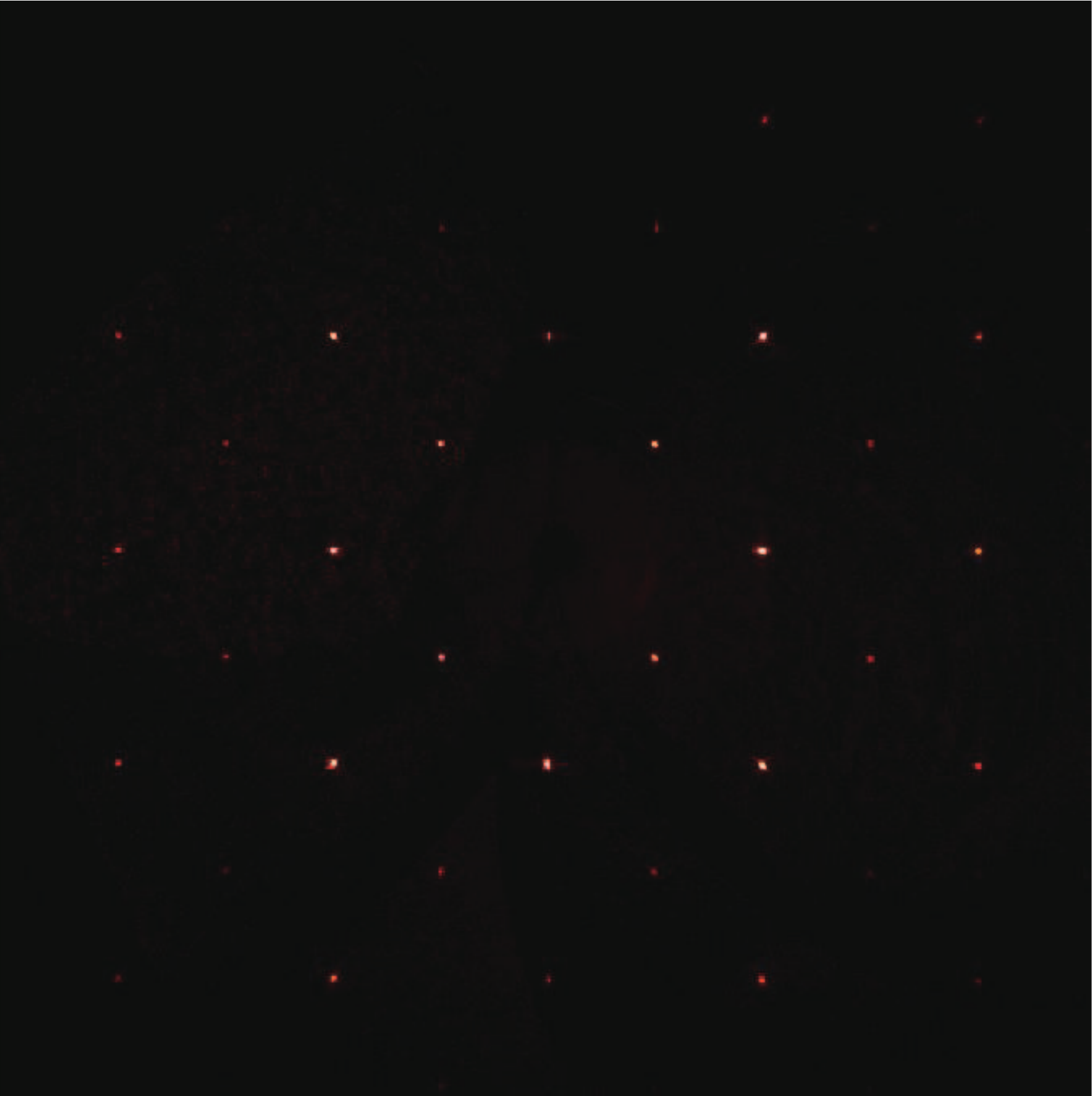}
\caption{The precession image in (h,k,0) plane for the SC-XRD measurement. All the peaks can be indexed by the I 4/m structure as shown in Table I.}
\end{figure}

\begin{table}
\centering
\begin{tabular}{ll}
\hline
\hline
Bond precision: & As-Fe = 0.0011 \AA \\
Wavelength: & 0.71073 \AA \\
Cell: & a = 3.8676(7) \AA, c = 12.287(4) \AA \\
	&$\alpha$ = 90$^{\circ}$, $\beta$ = 90$^{\circ}$, $\gamma$ = 90$^{\circ}$\\
Space group & I 4/m\\
Sum formula & As4 F0 Fe4 La0.68 Na1.32 O0 Ta0 \\
Mu(mm$^-1$) & 29.431\\
F000 &289.0\\
h, k, l max & 5, 5, 16\\
Nref & 127\\
R (reflections)& 0.0427(124)\\
wR2 (reflections)& 0.1302(127) \\
S& 1.252\\
Npar& 9\\
\hline
\hline
\end{tabular}
\caption{Parameters of La$_{0.34}$Na$_{0.66}$Fe$_2$As$_2$ at room temperature from SC-XRD \cite{supp}.}
\end{table}

Figure 2 shows the precession image in (h,k,0) plane for the SC-XRD measurement. Only well defined dots are observed, suggesting the high-quality of the sample. Table 1 shows the refinement results of the SC-XRD measurement, where the lattice parameters a and c are similar to those of La$_{0.5-x}$Na$_{0.5+x}$Fe$_2$As$_2$ \cite{YanJQ15,IyoA18}. There are 127 peaks that have been observed with the maximum (H,K,L) = (5,5,16). The ratio between La and Na is 0.34:0.66, which is close to that determined by EDX (0.39:0.61) and ICP (0.27:0.73). Considering the uncertainties in the EDX and ICP measurements due to some reasons that have been discussed previously \cite{YanJQ15}, such as the contamination from the flux and the reaction with water, the Na doping level is taken as $x$ = 0.16 according to the SC-XRD results as done previously for the $x$ = 0.1 single crystal \cite{YanJQ15}. This doping level is also consistent with results from other measurements as shown later.

Figure 3(a) shows the temperature dependence of normalized resistance $R_N$ = $R(T)/R(300 K)$, which confirms that $T_c$ is about 27 K. In La$_{0.4}$Na$_{0.6}$Fe$_2$As$_2$ \cite{YanJQ15}, a small upturn in $R(T)$ is found at $T_N$, whereas in our case, a broad hump feature can be seen. This is similar to those observed in La$_{0.5-x}$Na$_{0.5+x}$Fe$_2$As$_2$ for $x \geq$ 0.15 \cite{IyoA18}, Ba$_{1-x}$K$_x$Fe$_2$As$_2$ and Ba$_{1-x}$Na$_x$Fe$_2$As$_2$ around optimal doping level \cite{ShenB11,AswarthamS12,OhgushiK12}. The temperature dependence of $dR_N/dT$ shows a small dip at 106 K, which suggests the AF and structural transitions happen at the same temperature as in Ba$_{1-x}$K$_x$Fe$_2$As$_2$ \cite{ShenB11}.

\begin{figure}[tbp]
\includegraphics[width=\columnwidth]{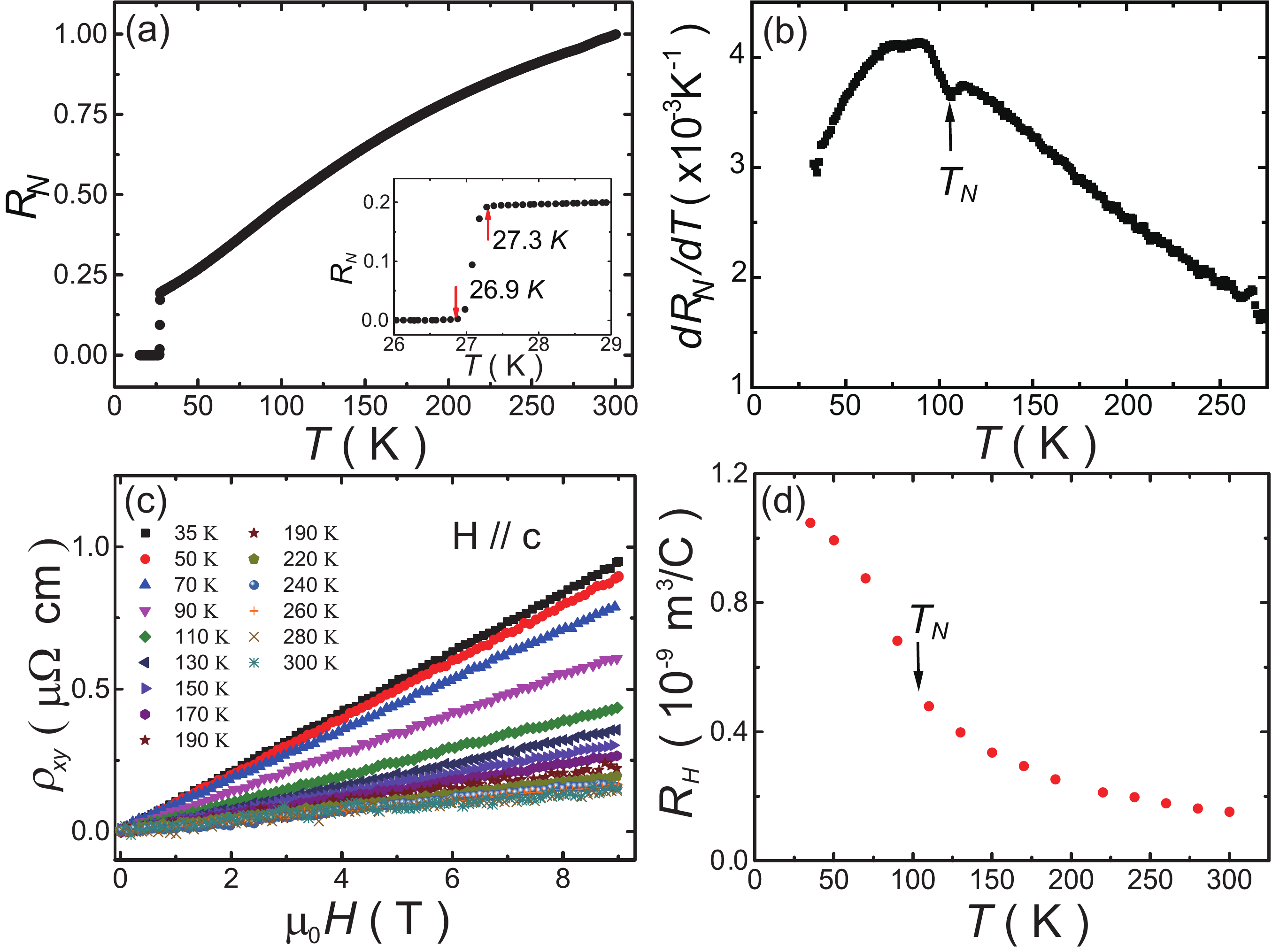}
\caption{(a) Temperature dependence of the normalized resistivity $R_N$ = $R(T)/R(300 K)$. The inset shows the data around the superconducting transition. (b) Temperature dependence of $dR_N/dT$. (c) Field dependence of Hall resistivity $\rho_{xy}$ at various temperatures with $H //$ c. (d) Temperature dependence of the Hall coefficient $R_H$.}
\end{figure}

Figure 3(c) shows the field dependence of the Hall resistivity $\rho_{xy}$ at different temperatures for $H//c$, which all show linear field dependence with positive slopes, confirming that the sample is hole-doped. Accordingly, we derive the temperature dependence of the Hall coefficient $R_H$ as shown in Fig. 3(d). The quick increase of $R_H$ below 100 K with decreasing temperature is consistent with the establishment of the AF order below $T_N$ \cite{ShenB11,AswarthamS12,OhgushiK12,LiuY14}. The values of $R_H$ are also similar to those of nearly optimally hole-doped "122" materials \cite{ShenB11,AswarthamS12,OhgushiK12,LiuY14}.

\begin{figure}[tbp]
\includegraphics[width=\columnwidth]{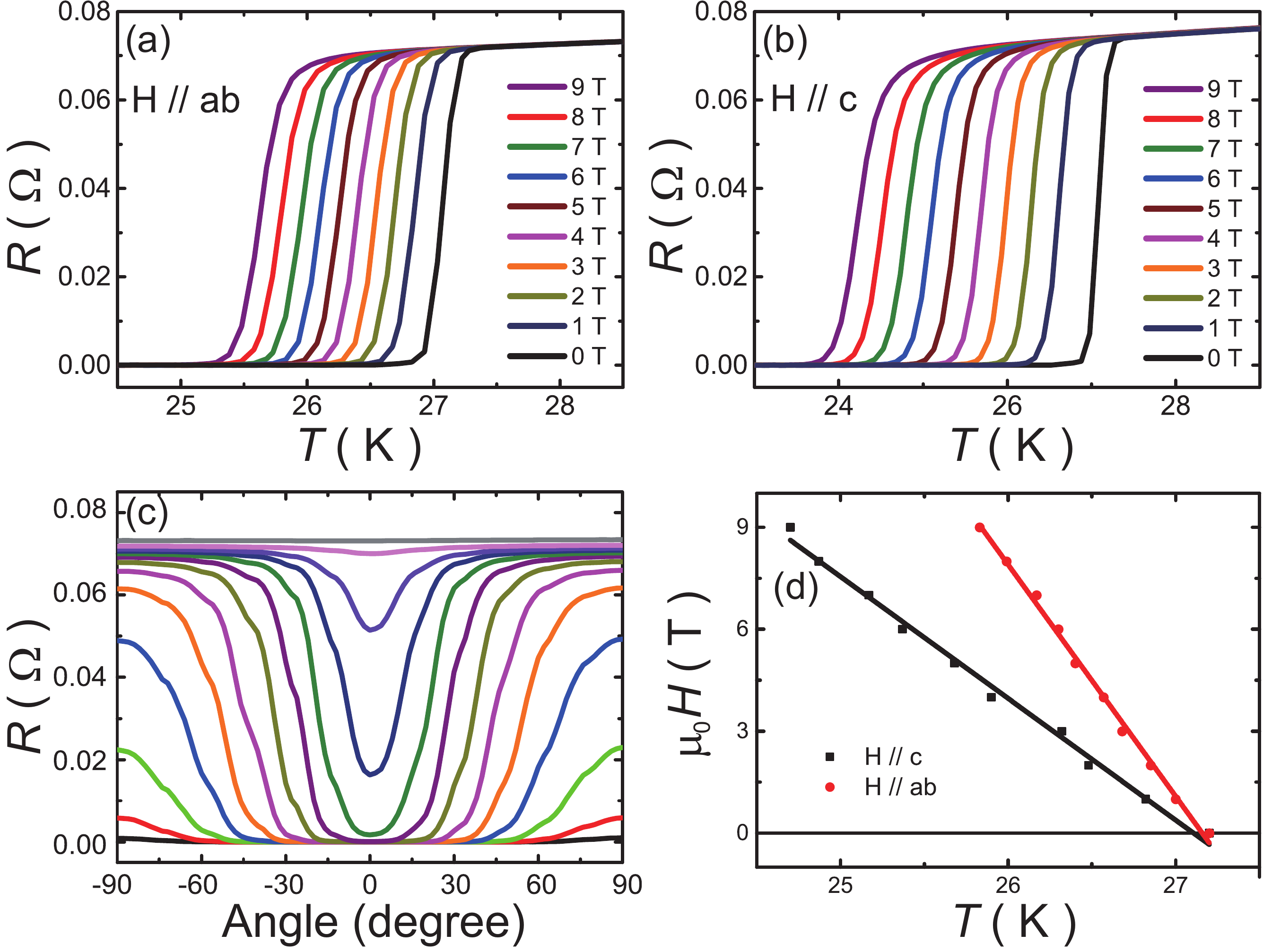}
\caption{(a) \& (b) Suppression of the superconductivity under magnetic fields for $H//ab$ and $H//c$, respectively. (c) Angle dependence of resistance under 9 Tesla from 23.6 K to 25.6 K with 0.2 K step, 26 K and 27 K, for lines from bottom to top. The zero and $\pm$ 90 degrees correspond to $H//c$ and $H//ab$, respectively. (d) $T_c$ for the field along c axis (black squares) and within the ab plane (red circles). The straight lines are linearly fitted results.}
\end{figure}

Fig. 4(a) and Fig. 4(b) shows the temperature dependence of the resistance for various magnetic fields applied within the ab plane and along the c axis, respectively. The broadening of the superconducting transition under field is not obvious. The suppression of $T_c$ for the field along the c axis is larger than that for the field within the ab plane. This anisotropy can be further confirmed by checking the angle dependence of the resistance with the field rotating between the ab plane and c axis, as shown in Fig. 4(c). Figure 4(d) shows the change of $T_c$ under field, which gives $dH_{c2}^{ab}(T) / dT \approx$ 6.8 T/K and $dH_{c2}^{c}(T) / dT \approx$ 3.6 T/K with $H_{c2}^{ab}$ and $H_{c2}^{c}$ representing the upper critical fields within the ab plane and along the c axis, respectively. The value of the zero-temperature upper critical field can be estimated by Werthamer-Helfand-Hohenberg formula $H_{c2}(0)=-0.693T_c(dH_{c2}/dT)|_{T_c}$. Taking  $T_c$ = 27 K and the values of upper critical fields are $H_{c2}^{ab}(0)\approx 128 T$, $H_{c2}^{c}(0)\approx 67 T$. The superconducting anisotropy $\gamma = H_{c2}^{ab}/H_{c2}^{c}$ is about 1.9, which is close to those in hole-doped "122" materials \cite{NiN08,WangZS08,SunDL09,HaberkornN11}.

Figure 5 shows the phase diagram of La$_{0.5-x}$Na$_{0.5+x}$Fe$_2$As$_2$ by summarizing the previous works on both polycrystals and single crystals \cite{YanJQ15,IyoA18}. We have measured several samples from different batches and the values of both $T_N$ and $T_c$ show no change. Therefore, our sample is near boundary where the AF order disappears and the superconductivity appears. It should be noted that the doping level $x$ in polycrystalline samples is nominal \cite{IyoA18}, while those in single crystals are determined by SC-XRD \cite{YanJQ15}.

\section{Discussions}

Our results provide the first example of growing superconducting single crystal of hole-doped La$_{0.5-x}$Na$_{0.5+x}$Fe$_2$As$_2$ with $x$ = 0.16. In the previous report \cite{YanJQ15}, the growth of the $x$ = 0.1 single crystal is due to the use of Al$_2$O$_3$ crucible in growing LaFeAsO. While Na is from the NaAs flux, varying its percentage in the mixture does not change $x$ in the final product. It has been suggested that the reaction between the NaAsO$_2$ formed in the growing process \cite{YanJQ11} and the Al$_2$O$_3$ crucible may consume the oxygen and finally result in the growth of the La$_{0.4}$Na$_{0.6}$Fe$_2$As$_2$ sample. In our case, Al$_2$O$_3$ crucible was used in the pretreatment of the LaFeAsO$_{1-y}$F$_y$ pellets, where there is no reaction between these two since the pellets remained the same shape before and after the treatment, and the crucibles remained clear except some little dark powder left in the touched areas between the pellets and the crucibles. In the final growing process, the mixture was put into the Ta tube without the Al$_2$O$_3$ crucible. Depending on the original fluorine content $y$ in the pellets, LaFeAsO$_{1-y}$F$_y$ and La$_{0.34}$Na$_{0.66}$Fe$_2$As$_2$ single crystals were grown when $x$ $\leq$ 0.5 and $x$ $\geq$ 0.8, respectively \cite{GuY18}. Considering the large amount of NaAs and NaF used, the La$_{0.5-x}$Na$_{0.5+x}$Fe$_2$As$_2$ should be the product from the oxygen-poor or oxygen-free environment. Moreover, it is most likely that the presence of NaF enables more Na doping into the samples. Whether higher Na doping level can be achieved with the change of the ratio between NaAs and NaF needs to be further studied.

The properties of La$_{0.34}$Na$_{0.66}$Fe$_2$As$_2$, such as the coexistence of magnetism and superconductivity, the value and temperature dependence of the Hall coefficient and the superconducting anisotropy, are similar to those of nearly optimally hole-doped "122" materials, as shown in the previous section. The coexistence of superconductivity and magnetism is very common in iron-based superconductors. Especially, what we observed in Fig. 3(b) and 3(d) are very similar to those in Ba$_{1-x}$K$_x$Fe$_2$As$_2$ \cite{ShenB11,AswarthamS12,OhgushiK12,LiuY14}, suggesting that this coexistence in our sample is intrinsic. Near optimal doping level, the effect of magnetic transition on the
sample is very weak so one can only observe it in $dR/dT$, as also shown in Ba$_{1-x}$K$_x$Fe$_2$As$_2$ \cite{ShenB11,AswarthamS12,OhgushiK12}. The dip in Fig. 3(b) is sharp enough to conclude that the La/Na content should be homogeneous, or one would expect a broad crossoverlike
feature. In polycrystal samples \cite{IyoA18}, the resistivity is an averaged results from different directions and it may not be
sensitive enough to see the dip in $dR/dT$.

It is surprising that the value of $T_c$ in our sample ($x$ = 0.16) is the same as the maximum value in polycrystalline samples for $x$ at 0.3. This difference may be due to different ways of determining the Na content, but the presence of $T_N$ in the $x$ = 0.16 single crystal suggests that it is not the optimally doped sample. Therefore, it is possible that the $T_c$ of single-crystal samples may be further enhanced, or the La$_{0.5-x}$Na$_{0.5+x}$Fe$_2$As$_2$ system does not have the typical superconducting dome found in most iron-based superconductors. Moreover, this may also be related to the $C_4$ magnetic phase that has been observed in many hole-doped "122" families \cite{AvciS14,BohmerAE15,WangL16,TaddeiKM16,TaddeiKM17}, since if it presents, the $T_c$ in our sample should not be the maximum in the
La$_{0.5-x}$Na$_{0.5+x}$Fe$_2$As$_2$ system. In any case, the successful growth of superconducting single crystals samples provides a new platform to investigate these interesting questions in iron-based superconductors.

\begin{figure}[tbp]
\includegraphics[width=\columnwidth]{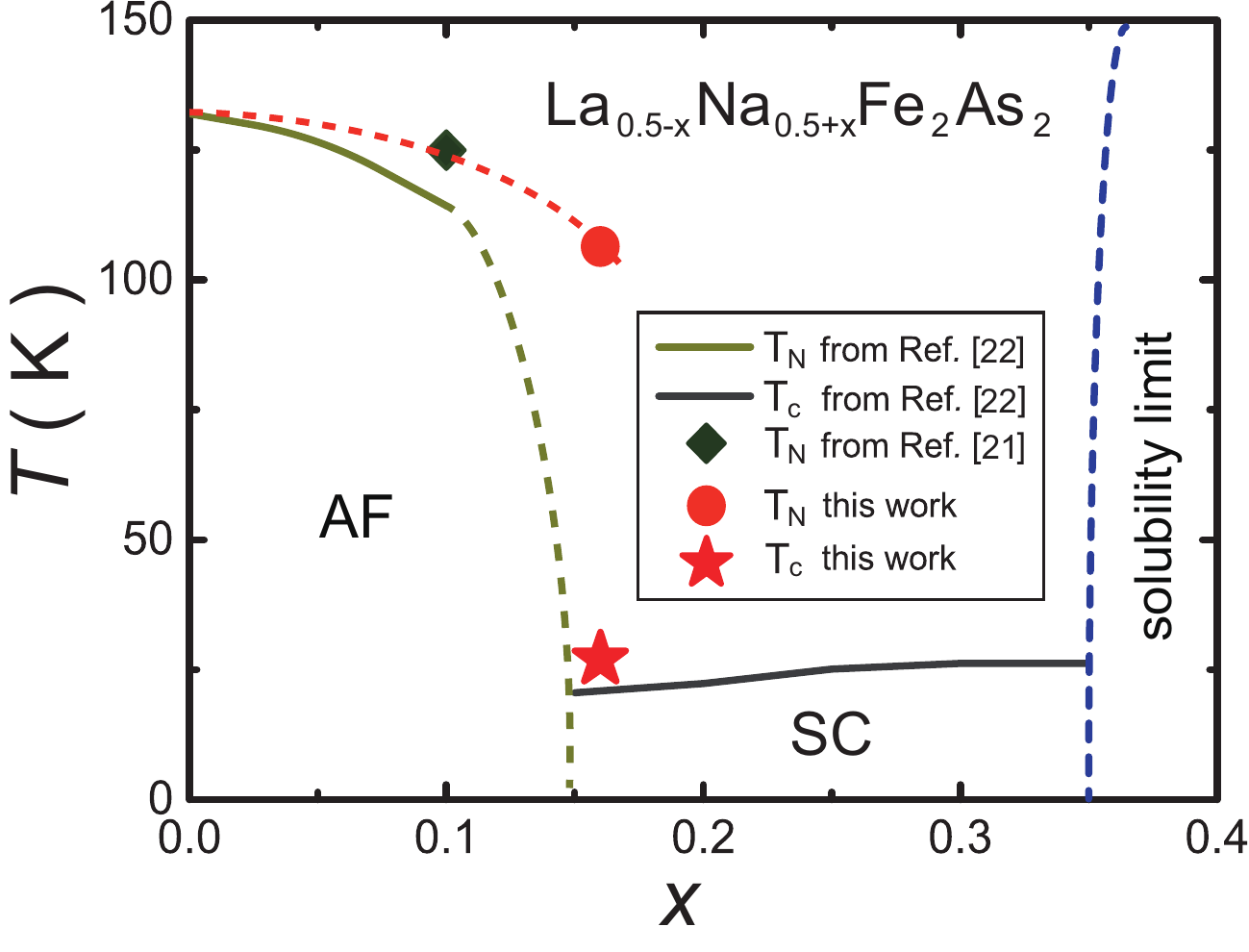}
\caption{Phase diagram of La$_{0.5-x}$Na$_{0.5+x}$Fe$_2$As$_2$. The antiferromagnetic (AF) and superconducting (SC) regimes are from measurements on polycrystalline samples \cite{IyoA18}. The diamond represents $T_N$ of the $x$ = 0.1 single crystal \cite{YanJQ15}. The circle and star represents $T_N$ and $T_c$ of the $x$ = 0.16 single crystal studied here, respectively. }
\end{figure}

\section{Conclusions}
We have successfully grown the superconducting La$_{0.34}$Na$_{0.66}$Fe$_2$As$_2$ single crystals by the flux method for the first time, which were accidentally obtained in growing LaFeAsO$_{1-x}$F$_x$ single crystals. The comparison between previous reports suggest that the ratio of NaAs/NaF may be important in controlling the doping level. The normal-state and superconducting properties of La$_{0.34}$Na$_{0.66}$Fe$_2$As$_2$ show that its doping level is slightly lower than optimal doping with $T_c$ = 27 K and $T_N = T_s$ = 106 K. Our results suggest that the La$_{0.5-x}$Na$_{0.5+x}$Fe$_2$As$_2$ system may be a new platform to study the antiferromagnetism and superconductivity in iron-based superconductors.

\ack
This work is supported by the Ministry of Science and Technology of China (No. 2017YFA0302900, No. 2016YFA0300502, No. 2017YFA0303103, No. 2016YFA0300604, No. 2015CB921300), the National Natural Science Foundation of China (No. 11674406, No. 11374011, No. 11774399, No. 11474330), the Strategic Priority Research Program (B) of the Chinese Academy of Sciences (XDB07020300, XDB07020200, XDB07020100, QYZDB-SSW-SLH043), and China Academy of Engineering Physics (No. 2015AB03). H. L. is supported by the Youth Innovation Promotion Association of CAS.

\providecommand{\newblock}{}

\end{document}